\documentclass{revtex4}
\usepackage{amssymb,amsfonts,amsmath,amsthm,mathtext,braket}

\begin{document}

\title{
On quantum channels generated by covariant positive operator-valued measures on a locally compact group
}

\author{\firstname{G.G.}~\surname{Amosov}}
\email[E-mail: ]{gramos@mi-ras.ru}
\address{Steklov Mathematical Institute of Russian Academy of Sciences, ul. Gubkina 8, Moscow 119991, Russia}

\received{}

\begin{abstract} 
We introduce  positive operator-valued measure (POVM) generated by the projective unitary representation of  a direct product of locally compact Abelian group $G$ with its dual $\hat G$. The method is based upon the Pontryagin duality allowing to establish an isometrical isomorphism between the space of Hilbert-Schmidt operators in $L^2(G)$ and the Hilbert space $L^2(\hat G\times G)$.
Any such a measure determines a pair of hybrid (containing classical and quantum parts) quantum channels consisting of the measurement channel and the channel transmitting an initial quantum state to the ensemble of quantum states on the group. It is shown that the second channel can be called  a complementary channel to the measurement channel.
\end{abstract}

\keywords{covariant positive operator-valued measure, projective unitary representation, locally compact Abelian group, Pontryagin duality} 

\maketitle

\section{Introduction}

Covariant positive operator-valued measures are often found in quantum information theory in different contexts \cite{Holevo}. At first, the task of their construction is of independent interest \cite{Heinosaari, Decker}.
Then, they define measurement channels for which the task of calculating capacity is set \cite{Holevo2, Holevo3} and noncommutative operator graphs that play an important role in the theory of quantum error correcting codes \cite{Amo, AMP2, AMP3, AMP4}. It should also be noted the use of operator-valued measures in quantum control problems \cite{Pechen_2015, Pechen_2018}.

To construct covariant operator-valued measures, it is natural to consider orbits of projective unitary representations of locally compact groups. At the same time, various problems arise related to the definition of an operator-valued measure of subsets of the group having infinite Haar measures \cite{Holevo}. Recently, new ideas have emerged for using the Pontryagin duality principle to correctly determine the integral over the orbits of the representation of locally compact groups. At the same time, a model was constructed that uniformly describes various types of quantum tomography in both finite-dimensional and infinite-dimensional Hilbert spaces \cite{tomogram}.

It is known that the process of measuring a quantum state leads to the emergence of an ensemble of posteriori quantum states. An important task is to estimate the amount of information contained in such an ensemble \cite{Holevo4}. Considering both the measurement channel and the ensemble of quantum states arising after the measurement simultaneously leads to the concept of a hybrid system. For ordinary channels acting on spaces of all states in a fixed Hilbert space, the concept of a complementary channel is defined. If the initial channel is obtained by averaging over the environment, then the complementary channel is given by averaging over the output system and sets the state of the environment \cite{Holevo5}. For hybrid systems, a similar approach was proposed in \cite{AMP} using ideas of \cite{Shirokov}.

In the present paper we establish an isometrical isomorphism between the space of Hilbert-Schmidt operators and $L^2(\hat G\times G,\hat \mu \times \mu)$, where $G$ and $\hat G$ are locally compact Abelian group and its dual. The Haar measures $\hat \mu $ on $\hat G$ and $\mu $ on $G$ are supposed to be connected by the Pontryagin duality. This technique allows us to prove that the orbits of projective unitary representation of $\mathfrak {G}=\hat G\times G$ generate the covariant positive operator-valued measure. Using the constructed covariant POVM we define two quantum channels. The first one is the measurement channel mapping an initial quantum state to the probability distribution on $\mathfrak {G}$. The second channel transmit an initial state to the ensemble of quantum states on $\mathfrak {G}$. Both the channels are hybrid because they have quantum and classical parts. It is shown that these two channels can be called complementary in relation to each other.

\section{Covariant operator-valued measures}

Let $X$ be a measurable space and $\mathfrak {B}(X)$ be the set of measurable subspaces $B\subset X$. Denote $B(H)$ and $B(H)_+$ the algebra of all bounded operators and the cone of positive operators in a Hilbert space $H$. A map $\mathfrak {M}:\mathfrak {B}(X)\to B(H)_+$ is said to be a positive operator-valued measure (POVM) iff
$$
\mathfrak {M}(\cup _jB_j)=\sum \limits _j\mathfrak {M}(B_j)\ \text {whenever}\  B_j\cap B_k=\emptyset ,\ j\neq k,\ B_j\in \mathfrak {B}(X);
$$
\begin{equation}\label{usl}
\mathfrak {M}(\emptyset )=0,\ \mathfrak {M}(X)={\rm I}\  \text {(the identity operator)}.
\end{equation}

Let $X=\mathfrak G$ be a locally compact Abelian group with the Haar measure $\nu $. 
Consider a projective unitary representation
$g\to U_g$ of the group $\mathfrak G$ in a Hilbert space $H$. A POVM $\mathfrak M$ is said to be covariant with respect to $g\to U_g$ iff
$$
U_g\mathfrak {M}(B)U_g^*=\mathfrak {M}(B+g),\ B\in \mathfrak {B}(\mathfrak {G}),\ g\in \mathfrak {G}.
$$
Given $M\in B(H)_+$ it is naturally to construct a covariant POVM on $\mathfrak {G}$ by means of the formula
\begin{equation}\label{form}
\mathfrak{M}(B)=\int \limits _{B}U_gMU_g^*d\nu (g),\ B\in \mathfrak {B}(\mathfrak {G}).
\end{equation}
To define correctly a measure in the form (\ref{form}) we should satisfy the conditions (\ref {usl}). Since $\mathfrak {G}$ is locally compact only the integral in (\ref {form}) may not exist for infinite subsets $B=\cup B_j\subset \mathfrak {G},\ B_j\in \mathfrak {B}(\mathfrak {G})$, in particular, for $B=\mathfrak{G}$. If it still exists it is unclear whether $\mathfrak {M}(\mathfrak {G})={\rm I}$.

Given a vector $\psi \in H$ denote $H_{\psi}\subset H$ the linear closure of $\{U_g\psi,\ g\in \mathfrak {G}\}$.

{\bf Proposition.} {\it Suppose that (\ref {form}) determines POVM on $\mathfrak {G}$ for $M=\ket {\psi}\bra {\psi}$.
Then, the map $T:H_{\psi }\to L^2(\mathfrak {G},\nu )$ defined by the formula
$$
(T\xi)(g)=\braket {U_g\psi,\xi},\ \xi \in H_{\psi},
$$
establishes an isometrical isomorphism.}

Proof.

$$
||T\xi ||^2=\int \limits _{\mathfrak G}|\braket{U_g\psi,\xi }|^2d\nu (g)=\int \limits _{\mathfrak G}{\rm Tr}(U_g\ket {\psi}\bra {\psi}U_g^*\ket {\xi}\bra {\xi})d\nu (g)=
$$
$$
\int \limits _{\mathfrak G}\braket {\xi,d\mathfrak {M}(g)\xi }=||\xi ||^2.
$$

$\Box$

\section {Covariant measure on a direct product $\hat G\times G$}

In the following we will need techniques of abstract harmonic analysis \cite{Rudin}.
Suppose that $G$ is a locally compact Abelian group. Then, one can define the Haar measure $\nu$ on the algebra $\mathfrak {B}(G)$ of subsets $B\subset G$ including compact $B$ which is invariant with respect to the action of $G$, $\nu (B+g)=\nu (B),\ g\in G,\ B\in \mathfrak {B}(G)$. The Haar measure $\nu $ is know to be unique up to multiplication by a positive constant. The homomorphism $\chi :G\to {\mathbb T}=\{z\in {\mathbb C}:\ |z|=1\}$ is said to be a character of $G$. The set of all characters forms the locally compact group $\hat G$ know as a dual to $G$. Taking $f\in L^1(G,\nu )$ it is possible to define the Fourier transform
\begin{equation}\label{Pontr}
{\mathcal F}(f)(\chi)=\int \limits _G\overline {\chi (g)}f(g)d\nu (g).
\end{equation}
Due to the Pontryagin duality \cite {Pontryagin} there exists the unique Haar measure $\hat \nu $ on $\hat G$ such that (\ref {Pontr}) can be extended to the isometrical isomorphism ${\mathcal F}:L^2(G,\nu )\to L^2(\hat G,\hat \nu )$ and the inverse Fourier transform ${\mathcal F}^{-1}$ is given by the formula
$$
{\mathcal F}^{-1}(\hat f)(g)=\int \limits _{\hat G}\chi (g)\hat f(\chi )d\hat \nu (\chi ),\ \hat f\in L^2(\hat G,\hat \nu)
$$
determining an isomorphism $\hat {\hat G}\cong G$.

Now put $\mathfrak {G}=\hat G\times G$ and $\mu =\hat \nu \times \nu $ and define a projective unitary representation of $\mathfrak {G}$ in the Hilbert  space $H=L^2(G,\nu )$ by the formula
\begin{equation}\label{represent}
[U_{\chi ,g}f](h)=\chi (h)f(h+g),\ f\in H,
\end{equation}
such that
$$
U_{\chi ,g}U_{\chi ',g'}=\chi (g')U_{\chi \chi ',g+g'},\ \chi ,\chi '\in \hat G,\ g,g'\in G.
$$
Denote $\mathfrak {S}(H)$ and $\mathfrak {S}_2(H)$ the convex set of quantum states (positive unit trace operators) and the space of Hilbert-Schmidt operators in $H$ correspondingly.

{\bf Theorem 1.} {\it The map $T:\mathfrak{S}_2(H)\to \mathcal {H}=L^2(\mathfrak {G},\mu )$ determined by 
$$
(T\rho)(\chi ,g)={\rm Tr}(\rho U_{\chi ,g})
$$
establishes an isometrical isomorphism such that
\begin{equation}\label{Parseval}
||T\rho ||_{\mathcal H}^2={\rm Tr}(\rho ^*\rho ).
\end{equation}
The inverse map is given by the formula
$$
T^{-1}F =\int \limits _{\mathfrak {G}}F(\chi ,g)U_{\chi ,g}^*d\hat \mu (\chi )d\mu (g),
$$
where convergence is understood in a weak operator topology.
}

Proof.

Take an orthonormal basis $(\psi _j)$ in $H$. Then, given $\rho \in \mathfrak {S}_2(H)$ there is a collection of complex numbers $(c_{jk}),\ \sum \limits _{j,k}|c_{jk}|^2<+\infty ,$ such that
$$
\rho =\sum \limits _{j,k}c_{jk}\ket {\psi _j}\bra {\psi _k}.
$$
Following to the idea of \cite {tomogram} (Proposition 1) put
\begin{equation}\label{funct}
F_{jk}(\chi ,g)={\rm Tr}(\ket {\psi _j}\bra {\psi _k}U_{\chi ,g})=\braket {\psi_k,U_{\chi ,g}\psi_j}.
\end{equation} 
Then,
$$
F_{jk}(\chi ,g)=\int \limits _{G}\overline \psi _k(h)\chi (h)\psi _j(h+g)d\mu (h)=\int \limits _{\mathfrak {G}}\chi '(g)\chi (h)\hat \psi _j(\chi ')\overline \psi _k(h)d\hat \mu (\chi ')d\mu (h).
$$
Thus, $F_{jk}$ is the Fourier transform of the function $\hat \psi _j(\chi )\psi _k(g)\in \mathcal {H}=L^2(\mathfrak {G},\mu)$.
It implies that (\ref {funct}) form the orthonormal basis in $\mathcal H$. Hence,
$$
||T\rho ||_{\mathcal H}^2=||\sum \limits _{j,k}c_{jk}T(\ket {\psi _j}\bra {\psi _k})||^2=||\sum \limits _{j,k}c_{jk}F_{jk}||^2=
$$
$$
\sum \limits _{j,k}|c_{jk}|^2={\rm Tr}(\rho ^*\rho ).
$$
Given $F\in \mathcal H$ and $\psi ,\xi \in H$ we obtain
\begin{equation}\label{est}
\braket {\psi ,T^{-1}(F)\xi }_H=\int \limits _{\mathfrak G}F(\chi ,g)\braket {U_{\chi ,g}\psi ,\xi }d\hat \mu (\chi )d\mu (g).
\end{equation}
Since $F(\chi ,g),\braket {\psi ,U_{\chi ,g}\xi }\in {\mathcal H}$ the integral (\ref {est}) converges in a weak operator topology. 
Moreover, the following resolution takes place
$$
F(\chi ,g)=\sum \limits _{jk}c_{jk}F_{jk}(\chi ,g),\ \sum \limits _{j,k}|c_{jk}|^2<+\infty ,
$$
where $F_{jk}$ are defined by (\ref {funct}). It results in
$$
T^{-1}F=\sum \limits _{j,k}c_{jk}\ket {\psi _j}\bra {\psi _k}\in \mathfrak {S}_2(H).
$$

$\Box $

{\bf Corollary 1.} {\it Given $\rho \in \mathfrak {S}(H)$ the formula
\begin{equation}\label{F}
\mathfrak {M}(B)=\int \limits _BU_{\chi ,g}\rho U_{\chi ,g}^*d\hat \mu (\chi)d\mu (g)
\end{equation}
determines a covariant POVM such that
$$
\int \limits _{\mathfrak G}U_{\chi ,g}\rho U_{\chi ,g}^*d\hat \mu (\chi)d\mu (g)={\rm I}.
$$
}

Proof.

Given $\rho \in \mathfrak {S}_2(H)$ denote $F_{\rho }(\chi ,g)={\rm Tr}(\rho U_{\chi ,g})$. 
It follows from Theorem 1 that the Parseval identity (\ref {Parseval}) holds true. Therefore,
\begin{equation}\label{parseval}
\int \limits _{\hat G\times G}F_{\rho }(\chi ,g)\overline {F_{\sigma}(\chi ,g)}d\hat \mu (\chi )d\mu (g)={\rm Tr}(\rho \sigma ^*)
\end{equation}
for all $\rho ,\sigma \in \mathfrak {S}_2(H)$.
Let us take the inner product for (\ref {F}) with $\rho =\ket {\psi}\bra {\psi}$, then
$$
\braket {\xi,\mathfrak {M}(B)\eta}=\int \limits _B\braket {\xi,U_{\chi ,g}\ket {\psi}\bra {\psi}U_{\chi ,g}^*\eta}d\hat \mu (\chi)d\mu (g)=
$$
$$
\int \limits _BF_{\ket {\psi}\bra {\xi}}(\chi ,g)\overline {F_{\ket {\psi}\bra {\eta}}(\chi ,g)}d\hat \mu (\chi)d\mu (g),\ \xi,\eta \in H.
$$
It follows that $\mathfrak {M}$ is positive and countably additive. Substituting $B=\hat G\times G$ and taking into account (\ref {parseval}) we obtain
$$
\braket {\xi,\mathfrak {M}(\hat G\times G)\eta}={\rm Tr}(\ket {\psi}\bra {\xi}\cdot \ket {\eta}\bra {\psi})=\braket {\xi,\eta}
$$
Since any $\rho \in \mathfrak {S}(H)$ can be represented as a convex sum
$$
\rho =\sum \limits _j\pi _j\ket {\psi_j}\bra {\psi_j}
$$
with $\braket {\psi_j,\psi_k}=\delta _{jk}$ and $\pi _j\ge 0,\ \sum \limits _j\pi _j=1$, the result follows.

$\Box $

Given a unit vector $\psi \in L^2(G,\nu)$ define a linear map $V:L^2(G,\nu)\to L^2(\mathfrak {G},\mu )$ by the formula
\begin{equation}\label{map}
(V\xi)(\chi ,g)=\braket {U_{\chi ,g}\psi ,\xi},\ \xi \in L^2(G,\nu).
\end{equation}

{\bf Corollary 2.} {\it The map (\ref {map}) establishes an isometrical isomorphism and
$$
\mathfrak {N}(B)=V{\mathfrak M}(B)V^*,\ B\in \mathfrak {B}(\mathfrak {G})
$$
is POVM on $\mathfrak {G}$ acting in the Hilbert space $\mathcal {H}=L^2(\mathfrak {G},\mu)$.}

Proof.

It follows from Proposition that $V$ is an isometrical isomorphism between the linear closure $H_{\psi }$ of $\{U_{\chi ,g}\psi \}$ and $L^2(\mathfrak {G},\mu )$. It follows from Corollary 1 that any vector $\xi \in L^2(G,\nu)$ can be represented as
$$
\xi =\int \limits _{\mathfrak G}\braket {U_{\chi ,g}\psi , \xi}U_{\chi ,g}\psi d\hat \mu (\chi )d\mu (g),
$$
where $\psi $ is a fixed unit vector ($\rho =\ket {\psi }\bra {\psi }$).
Hence, $\braket {U_{\chi ,g}\psi ,\xi }=0$ for all $(\chi ,g)\in \mathfrak {G}$ implies $\xi =0$ and 
$H_{\psi }=L^2(G,\nu )$.

$\Box $

{\bf Example.} {\it $G=\mathbb R$.}

The dual group $\hat G=\mathbb R$. Let the standard position and momentum operators $q$ and $p$
acts in the Hilbert space $H=L^2(G,\nu )\equiv L^2({\mathbb R})$ by the formula
$$
(q\psi )(x)=x\psi (x),\ \psi \in D(q)=\{\psi :\ x\psi (x) \in H\}, 
$$
$$
(p\psi )(x)=-i\frac {d}{dx}\psi ,\ \psi \in D(p)=\{\psi :\ \psi '(x) \in H\}
$$
Then, (\ref {represent}) can be represented as
$$
U_{x,y}=e^{ixq}e^{iyp},\ (x,y)\in {\mathbb R}\times {\mathbb R}.
$$
Put $\alpha =\frac {-y+ix}{\sqrt 2}$ and $a=\frac {q+ip}{\sqrt 2}$, it results in
$$
U_{x,y}=e^{-\alpha ^2+\overline \alpha ^2}D(\alpha ),
$$
where 
\begin{equation*}\label{dis}
D(\alpha )=exp(\alpha a^{\dag }-\overline {\alpha }a)
\end{equation*}
is the displacement operator with the creation and annihilation operators defined by the formula 
$$
a^{\dag }=\frac {q-ip}{\sqrt 2},\ a=\frac {q+ip}{\sqrt 2}.
$$
Take a function
$$
\psi _0=\frac {1}{\pi ^{1/4}}e^{-\frac {x^2}{2}},
$$
then
\begin{equation}\label{alpha}
\psi _{\alpha }=D(\alpha )\ket {0}
\end{equation}
is a coherent state for $\alpha \in \mathbb C$ and the measure defined by Theorem 1 takes the form
$$
\mathfrak {M}(B)=\frac {1}{\pi }\int \limits _{B}\ket {\alpha }\bra {\alpha }d^2\alpha .
$$
Here $L^2(\mathfrak {G},\mu)$ is the Bargmann-Fock space consisting of entire functions $\psi (z)$ equipped with the inner product 
$$
\braket {\psi ,\xi }=\int \limits _{\mathbb C}e^{-|\alpha|^2}\overline \psi (\alpha)\xi (\alpha)d^2\alpha.
$$
The isomorphism (\ref {map}) is fulfilled by the rule
$$
(V\xi)(\alpha)=\braket {\psi _{\overline \alpha },\xi },
$$
where $\psi _{\alpha}$ are coherent states (\ref {alpha}).

\section{Application to quantum channels}

Denote $\Pi (\mathfrak {G})$ the set of all probability distributions on $\mathfrak {G}$. Following to \cite {AMP} we can define two quantum channels corresponding to POVM (\ref {F}). One is the measurement channel $\Phi :\mathfrak {S}(H)\to \Pi (\mathfrak {G})$ such that
\begin{equation}\label{ProBab}
\Phi (\rho )(B)={\rm Tr}(\rho \mathfrak {M}(B)),\ \rho \in \mathfrak {S}(H),\ B\in \mathfrak {B}(\mathfrak {G}).
\end{equation}
Suppose that $d\mathfrak {M}(\chi ,g)=U_{\chi ,g}\ket {\psi}\bra {\psi }U_{\chi ,g}^*d\hat \nu (\chi )d\nu (g)$. Then, 
$$
\Phi (\rho )(B)=\int \limits _{B}p_{\rho }(\chi ,g)d\mu (\chi ,g)
$$
where the density of probability distribution $p_{\rho }$ determined by (\ref {ProBab}) is given by the formula
\begin{equation}\label{MC}
p_{\rho }(\chi ,g)=\braket {\psi ,U_{\chi ,g}^*\rho U_{\chi ,g}\psi}.
\end{equation}
The value of (\ref{MC}) at each fixed point $(\chi ,g)$ gives the mean of the observable $U_{\chi ,g}\ket {\psi }\bra {\psi}U_{\chi ,g}^*$ in the state $\rho$. In Example $(\chi ,g)=\alpha $ and $U_{\chi ,g}\psi \equiv D(\alpha )\psi _0$ is a coherent state (\ref {alpha}). Thus, for our Example (\ref {MC}) becomes the Husimi function of $\rho $.

To define another channel let us consider the von Neumann algebra ${\mathcal M}=B(H)\otimes L^{\infty }(\mathfrak {G})$ and a completely positive map $\Psi ^*:\mathcal {M}\to B(H)$ determined by the formula
$$
{\rm Tr}(\rho \Psi ^*(T\otimes f))=\int \limits_{\mathfrak {G}}f(\chi ,g)\braket {\psi ,U_{\chi ,g}^*TU_{\chi ,g}\psi }\braket {\psi ,U_{\chi ,g}^*\rho U_{\chi ,g}\psi }d\hat \nu (\chi )d\nu (g).
$$
The pre-conjugate map $\Psi :\mathfrak{S}(H)\to \mathcal {M}_*$ is a quantum channel. Its action can be represented directly as follows
\begin{equation}\label{AC}
\Psi (\rho )(\chi ,g)=\braket {\psi ,U_{\chi ,g}^*\rho U_{\chi ,g}\psi}\ket {U_{\chi ,g}\psi }\bra {U_{\chi ,g}\psi},\ \rho \in \mathfrak {S}(H).
\end{equation}
The physical sense of (14) is the following. After the measurement the state of quantum system is determined by the ensemble $(\pi _{\chi ,g},\rho _{\chi ,g})$, where $\rho _{\chi ,g}=\ket {U_{\chi ,g}\psi }\bra {U_{\chi ,g}\psi}$ and $\pi _{\chi ,g}=\braket {\psi ,U_{\chi ,g}^*\rho U_{\chi ,g}\psi}$ is a probability distribution on $(\rho _{\chi ,g})$.

 Denote $S(\rho )=-{\rm Tr}(\rho \log \rho )$ the von Neumann entropy of a state $\rho \in \mathfrak {S}(H)$. There exists a natural extension of $S$ to ${\mathcal M}_*$ consisting of measurable functions $\rho _{\chi ,g}$ on $\mathfrak {G}$ with values in $\mathfrak {S}(H)$ satisfying the relation
$$
\int \limits _{\mathfrak {G}}\rho _{\chi ,g}d\hat \mu (\chi )d\mu (g)=1.
$$
It suffices to put
\begin{equation}\label{entropy}
S(\rho _{\chi ,g})=-\int \limits _{\mathfrak {G}}{\rm Tr}(\rho _{\chi ,g}\log \rho _{\chi ,g})d\hat \nu (\chi )d\nu (g).
\end{equation}
Formula (\ref {entropy}) can be considered also as an extension of the entropy of probability distribution $p \in \Pi (\mathfrak {G})$ defined by
$$
S(p)=-\int \limits _{\mathfrak {G}}p(\chi ,g)\log p(\chi ,g)d\hat \nu (\chi)d\nu (g)
$$
The following statement allows to call $\Psi $ to be complementary to the measurement channel $\Phi $. 

{\bf Theorem 2.} {\it There exists the isometrical embedding $W: H\to {\mathcal H}=H\otimes L^2(\mathfrak {G})$ such that
$$
d\Phi (\rho )(\chi ,g)=p_{\rho}d\mu (\chi ,g),\ p_{\rho}(\chi ,g)={\rm Tr}_H([W\rho W^*](\chi ,g)), 
$$
$$
\Psi (\rho )=\int \limits _{\mathfrak {G}}[W\rho W^*](\chi ,g)d\hat \nu (\chi )d\nu (g),\ \rho \in \mathfrak {S}(H),
$$
and
$$
S(\Phi (\ket {\xi }\bra {\xi }))=S(\Psi (\ket {\xi }\bra {\xi }))
$$
for all unit vectors $\xi \in H$.
}

Proof.

Let us define $W$ on the unit vector $\xi \in H$ by the formula
$$
[W\xi ](\chi ,g)=\braket {U_{\chi ,g}\psi ,\xi }U_{\chi ,g}\psi .
$$
Then,
$$
\braket {W\xi ,W\eta }_{\mathcal H}=\int \limits _{\mathfrak {G}}\braket {\xi ,U_{\chi  ,g}\psi}_H\braket {U_{\chi ,g}\psi ,\eta }_{H}d\hat \mu (\chi )d\mu (g)=\braket {\xi ,\eta }_H
$$
due to Corollary 2.

$\Box $

\section {Conclusion}

We consider the projective unitary representation (\ref {represent}) of the group $\mathfrak {G}$ being a direct product of locally compact Abelian group $G$ and its dual $\hat G$. Based upon the Pontryagin duality we have constructed the isometrical isomorphism between the space of Hilbert-Schmidt operators in $L^2(G)$ and $L^2(\mathfrak {G})$ (Theorem 1). It allows us to prove that orbits of our representation generate a positive operator-valued measure (\ref {F}). This measure, in turn, defines two hybrid quantum channels (\ref {MC}) and (\ref {AC}). These channels are shown to be complementary each to other (Theorem 2). 

\section*{Acknowledgment} This work is supported by Russian Science
Foundation under the grant No 19-11-00086.

\section *{Data Availability Statements} Data sharing not applicable to this article as no datasets were generated or analysed during the current study.

\end{document}